\documentclass[journal]{IEEEtran}
\usepackage{cite}
\usepackage[pdftex]{graphicx}
\usepackage{subcaption}
\usepackage{amsmath}
\usepackage{array}
\usepackage{hyperref}
\usepackage{url}
\begin{document}
\title{Transmission of signals in the 300~GHz band with a bit-error rate below $\textnormal{10}^{-9}$ using a soliton comb}
\author{Mantaro~Imamura,~Ryo~Sugano,~Ayaka~Yomoda,~Atsuro~Shirasaki,\\
    Koya~Tanikawa,~Soma~Kogure,~Shun~Fujii,~and~Takasumi~Tanabe
\thanks{MI, RS, AY, AS, KT, SK, and TT are with the Department of Electronics and Electrical Engineering, Faculty of Science and Technology, Keio University, Yokohama, Japan. e-mail: (takasumi@elec.keio.ac.jp).}
\thanks{SF is with the Department of Physics, Faculty of Science and Technology, Keio University, Yokohama, Japan.}}%
\maketitle
\begin{abstract}
To address the increasing demand for ultra-high-capacity wireless communication, terahertz (THz) frequencies near 300~GHz are attracting attention as a new spectral frontier. This work presents the first experimental demonstration of error-free (BER $< 1\times10^{-9}$) 10~Gbps transmission in the 300~GHz band using a soliton microcomb generated in an integrated silicon nitride (SiN) microring resonator. While many previous microcomb-based THz demonstrations have focused on coherent modulation formats and operation near the forward-error-correction (FEC) limit, this work investigates a simple intensity-modulation/direct-detection (IM-DD) on-off keying (OOK) architecture suitable for low-complexity THz links and fiber-wireless integrated systems. Although the experiment was conducted in a short back-to-back waveguide configuration, the generated THz wave enabled stable low-BER transmission without FEC or advanced offline signal processing. Analysis of the error-free threshold power indicates the feasibility of free-space transmission over several tens of meters with high-gain antennas and THz-band amplifiers. These results demonstrate the feasibility of robust low-complexity THz photonic links based on soliton microcombs for short-range fiber-wireless integrated systems.
\end{abstract}

\begin{IEEEkeywords}
terahertz-band communication, micro comb, soliton comb, photomixing.
\end{IEEEkeywords}

\section{Introduction}
\IEEEPARstart{H}{igh}-capacity wireless communication systems are needed to support the rapid growth of data traffic in next-generation networks. Currently, frequencies up to several gigahertz are mainly used for wireless communication; however, these bands are already becoming congested. To secure wider bandwidths, it is essential to utilize higher carrier frequencies. Consequently, terahertz (THz) frequencies, which are close to 300~GHz, have attracted considerable attention as a new spectral frontier for high-capacity wireless communication. The World Radiocommunication Conference 2019 (WRC-19) identified frequency windows within the 275 to 450~GHz range for fixed and land-mobile services~\cite{WRC-19}, and IEEE 802.15.3d has defined point-to-point channels within the range of 252.72 to 321.84~GHz~\cite{IEEE 802.15.3d}.

THz waves occupy the frequency region between radio and optical waves, thus inheriting the advantages of both—wide bandwidth and strong directionality. These features make them attractive for applications such as spectroscopy and sensing~\cite{THz applications}. For wireless communication applications, however, significant atmospheric attenuation and molecular absorption~\cite{Propagation loss} restrict the transmission distance, confining most current THz links to short-range environments such as data-center interconnects~\cite{Data center} and kiosk terminals~\cite{Kiosk}.

Various methods have been proposed for generating THz waves, which can be broadly classified into electrical up-conversion and photonic down-conversion. Electrical approaches allow for compact implementation, but seamless integration with optical-fiber networks and photonic carrier distribution can be challenging. In contrast, photonic approaches, especially photomixing two optical tones using a high-speed photodiode, such as a uni-traveling-carrier photodiode (UTC-PD)~\cite{UTC-PD} can provide stable carrier generation and compatibility with optical fiber communication systems~\cite{Elec vs Photo 1,Elec vs Photo 2,Elec vs Photo 3}. Recently, microresonator-based optical frequency combs (microcombs) have emerged as promising light sources for photomixing~\cite{THz Photonic signal processing 1}. A microcomb consists of multiple equidistant optical modes, and when all comb lines are phase-locked in a dissipative Kerr soliton (DKS) state, it provides high mutual coherence and low phase noise. The free spectral range (FSR) of the comb is inversely proportional to the resonator size, and CMOS-compatible silicon nitride (SiN) microrings can precisely realize FSRs close to 300~GHz~\cite{SiN}. These characteristics make soliton microcombs attractive photonic sources for THz carrier generation~\cite{WGM by soliton 1,WGM by soliton 2}.

Several studies have reported THz-band transmission based on soliton microcombs~\cite{THz communication by soliton 1,THz communication by soliton 2,THz communication by soliton 3,THz communication by soliton 4,THz communication by soliton 5,THz communication by soliton 6}. However, most previous demonstrations were limited to performance near the forward error correction (FEC) limit, typically requiring offline signal processing or FEC. While many previous studies have focused on high-capacity coherent transmission schemes, robust low-BER transmission (BER~$< 10^{-9}$) using simple intensity-modulation/direct-detection on-off keying (IM-DD/OOK) architectures remains attractive for low-complexity THz links and fiber-wireless integrated systems such as radio-over-fiber (RoF), yet demonstrations based on soliton microcombs have remained limited. Demonstrating stable low-BER operation without relying on offline processing or FEC is therefore important for simple and hardware-efficient THz photonic links, particularly for short-range applications such as data-center interconnects and on-board wireless connections.

In this work, we address this challenge. This study represents the first experimental demonstration of error-free (BER~$< 10^{-9}$) 10~Gbps transmission at 300~GHz using a soliton microcomb in a simple IM-DD/OOK architecture. While many previous microcomb-based THz demonstrations have focused on coherent modulation formats and near-FEC-limit operation, low-complexity THz links based on IM-DD/OOK architectures remain attractive for short-range applications that require simple hardware configurations and low-latency operation. Such architectures are compatible with fiber-wireless integrated systems, including radio-over-fiber (RoF) networks, where photonic generation and distribution of THz carriers can simplify front-end implementation. Moreover, OOK-based physical-layer (PHY) modes are already included in standards such as IEEE 802.15.3d for short-range THz communication systems. These considerations motivate the investigation of robust low-BER THz transmission using soliton microcombs in simple IM-DD/OOK configurations with minimal digital signal processing (DSP) requirements.

The remainder of this paper is organized as follows. Section~\ref{sec2} describes the generation and stabilization of the soliton comb. Section~\ref{sec3} details the experimental setup and transmission results in the 300~GHz band. Section~\ref{sec4} discusses the scalability toward free-space THz links, and Section~\ref{sec5} concludes this paper.

\section{Soliton comb generation and stabilization}\label{sec2}
\begin{figure}[tb]
    \begin{center}
        \includegraphics[width=80mm]{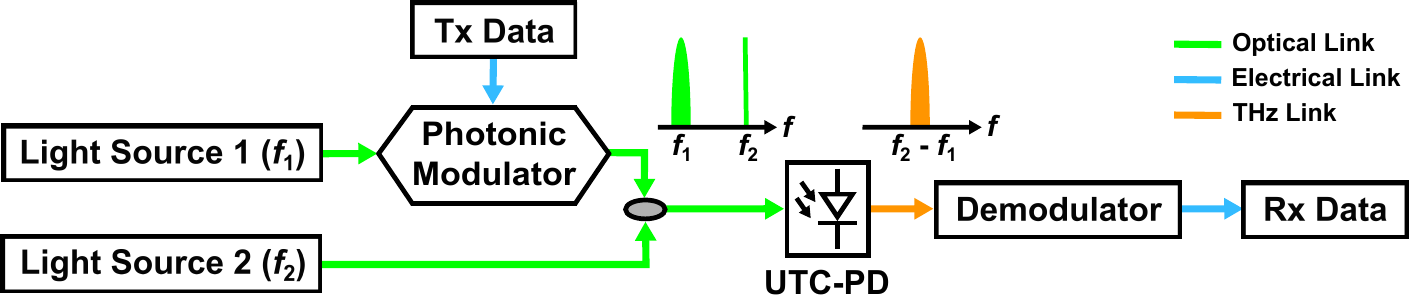}
    \end{center}
    \caption{THz band wireless communication by photomixing.}
    \label{fig:1}
\end{figure}

Figure~\ref{fig:1} shows THz-band wireless communication by photomixing. A beat signal at a frequency of $|f_\mathrm{1}-f_\mathrm{2}|$ is generated by mixing two optical signals with frequencies $f_\mathrm{1}$ and $f_\mathrm{2}$. A UTC-PD is used for photomixing. By modulating one of the optical carriers using a Mach–Zehnder modulator, the generated THz beat signal is also modulated.

In this work, a microresonator-based optical frequency comb (microcomb) was used as the optical source for photomixing. Microcombs consist of equidistant optical modes generated through nonlinear optical processes in a microresonator. Since the repetition frequency $f_\mathrm{rep}$ is inversely proportional to the resonator radius, compact SiN microring resonators can realize repetition frequencies in the THz range~\cite{SiN}. In particular, dissipative Kerr soliton (DKS) microcombs~\cite{WGM by soliton 1,WGM by soliton 2} provide coherent comb lines suitable for THz carrier generation.

The offset between the pump light frequency $f_\mathrm{pump}$ and the resonance frequency $f_\mathrm{res}$ is referred to as the detuning. Precise control of the pump-resonance detuning, enabling the transition from blue detuning ($f_\mathrm{pump}>f_\mathrm{res}$) to red detuning ($f_\mathrm{res}>f_\mathrm{pump}$), is essential for stable soliton generation~\cite{DKS}. In this work, the power-kicking method was employed to access the soliton state~\cite{Powerkick}.

\begin{figure*}[tbh]
    \begin{center}
        \includegraphics[width=0.7\linewidth]{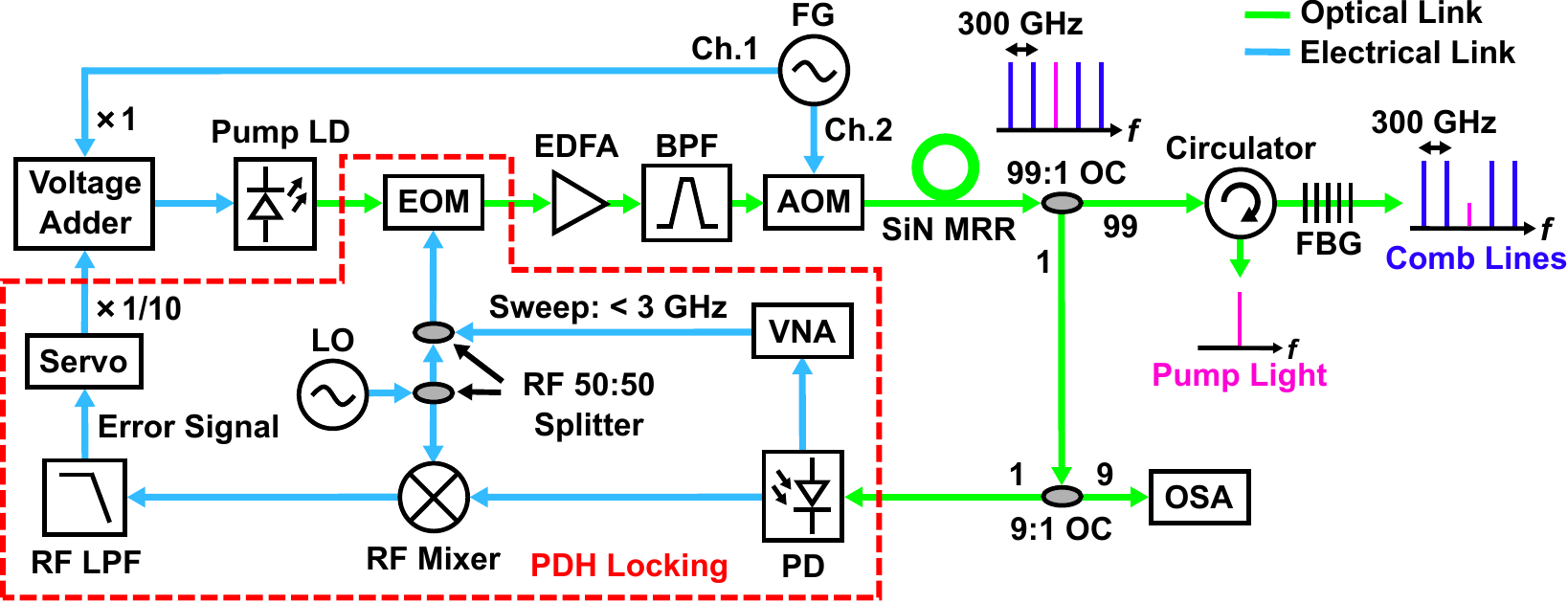}
    \end{center}
    \caption{Experimental setup for soliton comb generation. Pump~LD:~Pump~laser~diode, EOM:~Electro-optic~modulator, EDFA:~Erbium-doped~fiber~amplifier, BPF:~Band-pass~filter, AOM:~Acousto-optic~modulator, FG:~Function~generator, MRR:~Microring~resonator, OC:~Optical~coupler, FBG:~Fiber~bragg~grating, OSA:~Optical~spectrum~analyzer, PD:~Photodetector, VNA:~Vector~network~analyzer, LO:~Local~oscillator, RF~LPF:~Radio~frequency~low-pass~filter.}
    \label{fig:2}
\end{figure*}

\begin{figure}[tbh]
    \begin{center}
        \includegraphics[width=80mm]{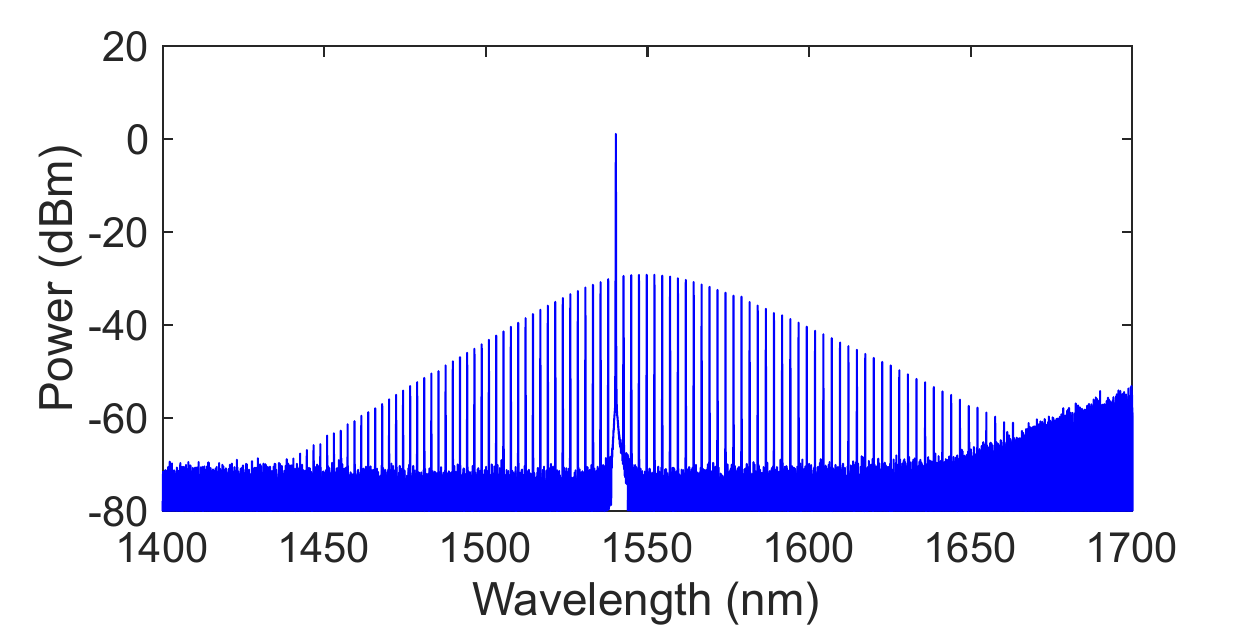}
    \end{center}
    \caption{Single soliton.}
    \label{fig:3}
\end{figure}

Figure~\ref{fig:2} shows the experimental setup for soliton comb generation and stabilization. The SiN microring resonator used in this study had a FSR of 300~GHz and a $Q$ factor of $2\times10^6$. Continuous-wave light at 1540.21~nm from a pump laser diode (Pump LD, santec, TSL-710) was amplified to 1~W using an erbium-doped fiber amplifier (EDFA). An acousto-optic modulator (AOM, Gooch\& Housego, T-M040-0.5C8J-3-F2P) was inserted before the resonator to enable power kicking. A function generator (FG, Keysight, 33600A) synchronized the AOM drive and the pump-frequency sweep. An optical band-pass filter (OBPF) was used to suppress amplified spontaneous emission from the EDFA. The output light from the resonator was separated into pump and comb components using a fiber Bragg grating (FBG). Figure~\ref{fig:3} shows the optical spectrum of the generated single-soliton state, exhibiting coherent comb lines with a spacing corresponding to the 300~GHz resonator free spectral range.

\begin{figure}[tbh]
    \begin{center}
        \includegraphics[width=80mm]{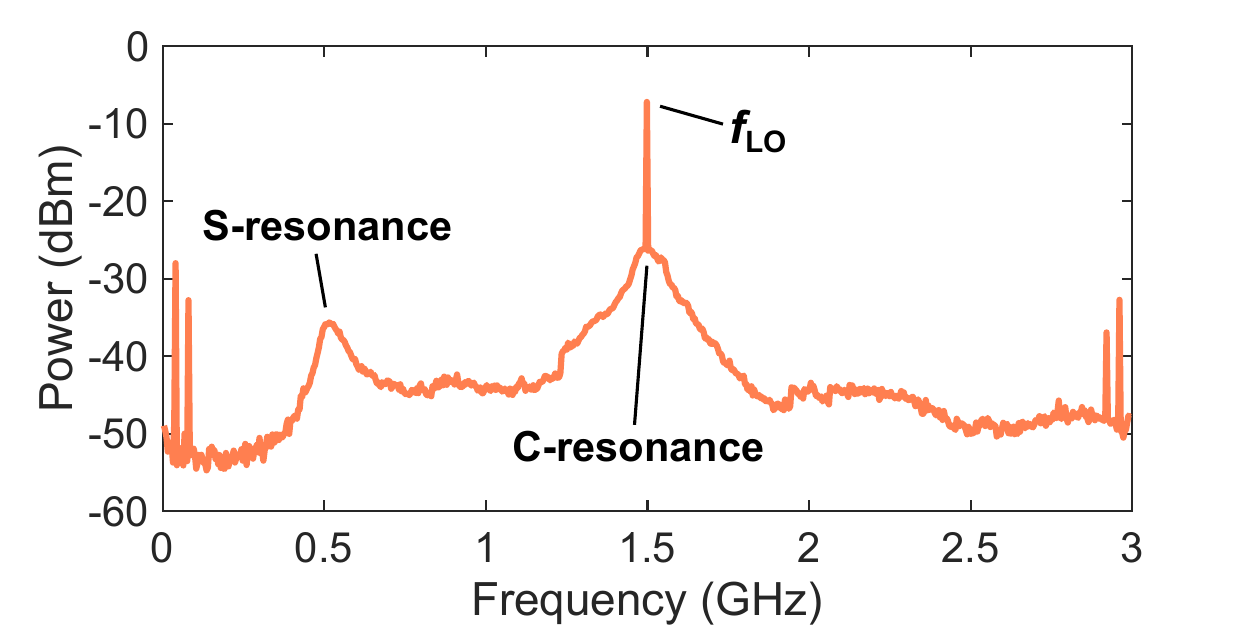}
    \end{center}
    \caption{Transfer function measured by the VNA.}
    \label{fig:4}
\end{figure}

In addition to stabilizing the resonator temperature, Pound–Drever–Hall (PDH) feedback control was applied to the pump laser to maintain a constant detuning~\cite{PDH}. A vector-network-analyzer-based (VNA-based) method was used to monitor the detuning~\cite{Resonance}. A vector network analyzer (VNA, Siglent, SSA 3032X-R) supplied a swept signal below 3~GHz, which was combined with a local oscillator (LO, Agilent, N9310A) signal at frequency $f_\mathrm{LO}$ and applied to an electro-optic modulator (EOM, Photline Technologies, MPZ-LN-20) to phase-modulate the pump light. Figure~\ref{fig:4} shows the measured transfer function. The 0.5~GHz peak corresponds to the S-resonance induced by the intracavity soliton, whereas the 1.5~GHz peak corresponds to the C-resonance and equals the detuning. The C-resonance peak was servo-controlled to match the LO frequency ($f_\mathrm{LO}=1.5013$~GHz). With combined control of the resonator temperature and pump frequency, the soliton state was maintained for more than ten hours.

\section{Transmission experiment in the 300 GHz}\label{sec3}
Although coherent quadrature amplitude modulation (QAM) transmission is an important future direction for microcomb-based THz systems~\cite{Bowen}, this study investigates a simple IM-DD/OOK architecture suitable for low-complexity THz links and fiber-wireless integrated systems. The goal is to demonstrate the feasibility of robust low-BER THz transmission rather than exhaustive telecom-grade system optimization. IM-DD is not only low-cost but also compatible with integrated photonic platforms such as Si photonics, and demonstrating stable low-BER transmission using THz waves generated from a soliton microcomb is important in itself.

\begin{figure*}[tbh]
    \begin{center}
        \includegraphics[width=0.8\linewidth]{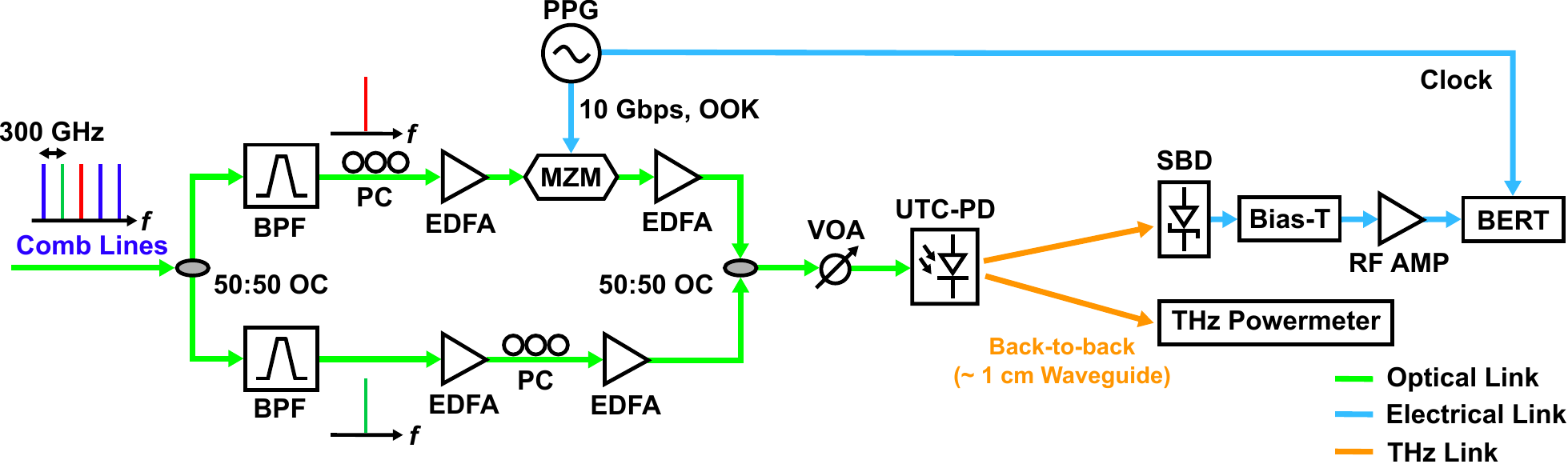}
    \end{center}
    \caption{Experimental setup for transmission in the 300 GHz. OC:~Optical~coupler, BPF:~Band-pass~filter, PC:~Polarization~controller, EDFA:~Erbium-doped~fiber~amplifier, MZM:~Mach-Zehnder~modulator, PPG:~Pulse~pattern~generator, VOA:~Variable~optical~attenuator, OSA:~Optical~spectrum~analyzer, UTC-PD:~Uni-traveling-carrier~photodiode, SBD:~Schottky~barrier~diode, RF~AMP:~Radio~frequency~amplifier, BERT:~Bit~error~rate~tester.}
    \label{fig:5}
\end{figure*}

Figure~\ref{fig:5} shows the experimental setup used for THz transmission experiments. Two adjacent comb lines (1552.14~nm and 1554.56~nm) were extracted from the generated single soliton using optical band-pass filters (BPFs). A pulse pattern generator (PPG, Anritsu, MP1761C) generated a 10-Gbps NRZ pseudo-random binary sequence signal with a word length of $2^7-1$. The output signal from the PPG was applied to a lithium-niobate Mach–Zehnder modulator (MZM, Photline Technologies, MXAN-LN-10) to modulate one comb line at 1552.14~nm. The polarization states of the optical signals were adjusted using polarization controllers (PCs), amplified by EDFAs, and combined using a 50:50 optical coupler. A variable optical attenuator (VOA) was used to adjust the optical power injected into the UTC-PD (NTT Innovative Devices, IOD-PMJ-13001). The maximum optical input power was 15.4~dBm, corresponding to a photocurrent of 7.0~mA. The generated 300~GHz signal was transmitted through a short back-to-back waveguide link of approximately 1~cm.

\begin{figure}[tbh]
    \begin{center}
        \includegraphics[width=80mm]{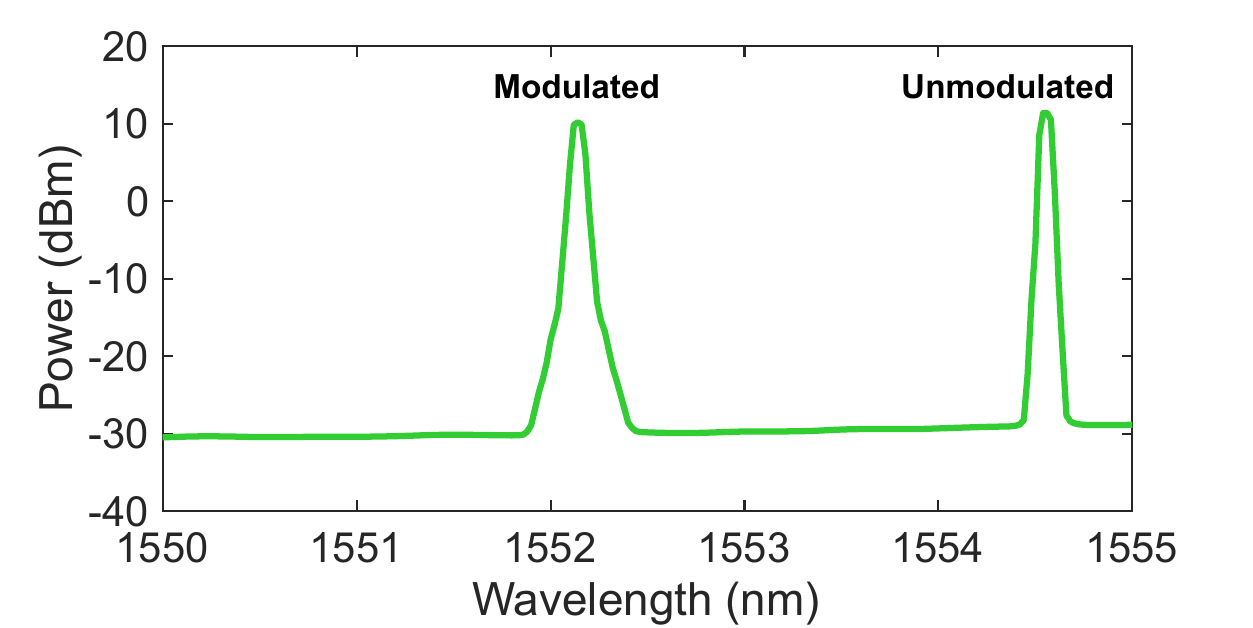}
    \end{center}
    \caption{UTC-PD input.}
    \label{fig:6}
\end{figure}

For BER measurements, the transmitted THz signal was detected using a Schottky barrier diode (SBD, VDI, WR3.4ZBD-F20). A bias-T removed the DC component, and a radio-frequency amplifier (RF AMP, Mini-Circuits, ZVA-213UWX+) amplified the AC component by 14~dB before detection by a bit-error-rate tester (BERT, Anritsu, MP1762C). The PPG and the BERT were synchronized using a common clock signal. Figure~\ref{fig:6} shows the optical spectrum of the UTC-PD input signal. The optical signal-to-noise ratio (OSNR) of the modulated comb line was measured to be 40.23~dB.

\begin{figure}[tbh]
    \begin{center}
        \includegraphics[width=80mm]{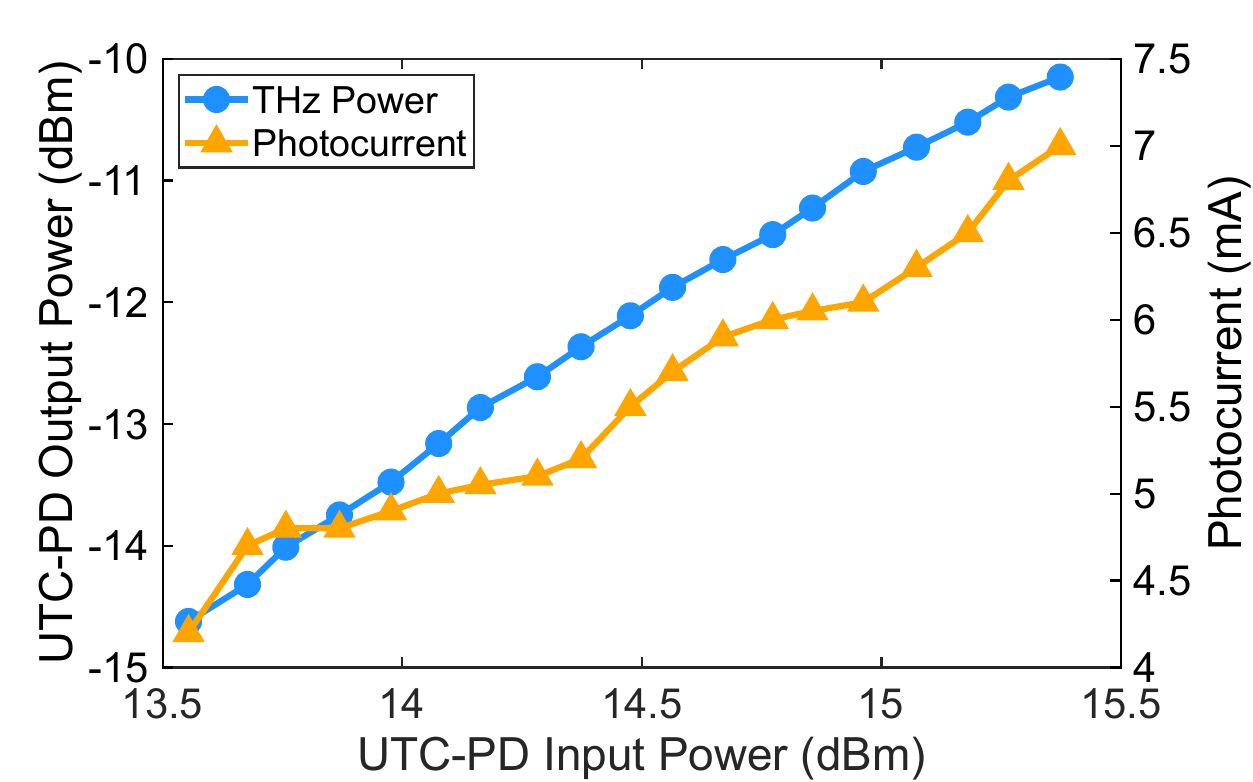}
    \end{center}
    \caption{UTC-PD output power and photocurrent.}
    \label{fig:7}
\end{figure}

Figure~\ref{fig:7} shows the photocurrent and UTC-PD output power as functions of the input optical power. Both the photocurrent and the generated THz power increased monotonically with increasing optical input power, and the UTC-PD output power exhibited an approximately linear dependence. At the maximum photocurrent of 7.0~mA, the UTC-PD output power reached $-10.15$~dBm.

\begin{figure}[tbh]
 \centering

 \begin{subfigure}{0.47\linewidth}
  \centering
  \includegraphics[width=\linewidth]{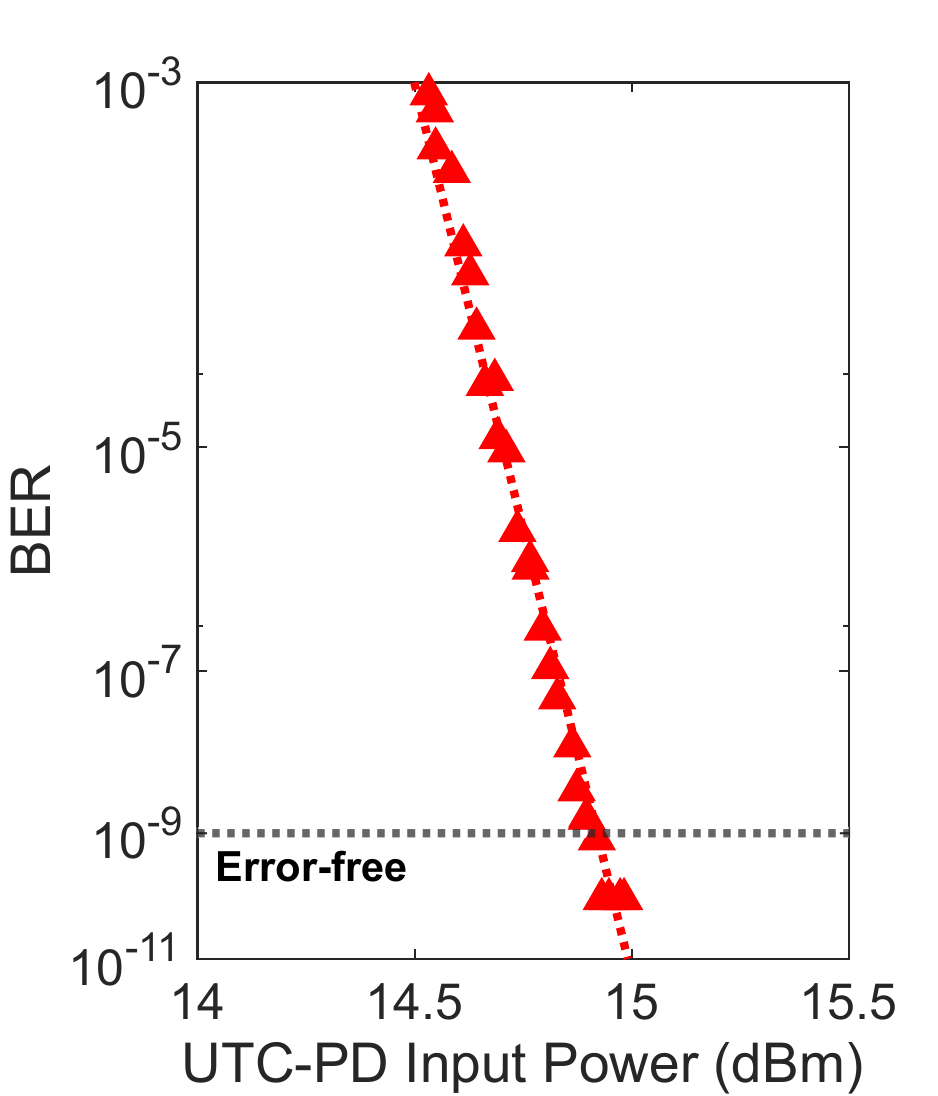}
  \caption{BER as a function of UTC-PD input optical power.}
  \label{fig:8a}
 \end{subfigure}\hfill
 \begin{subfigure}{0.47\linewidth}
  \centering
  \includegraphics[width=\linewidth]{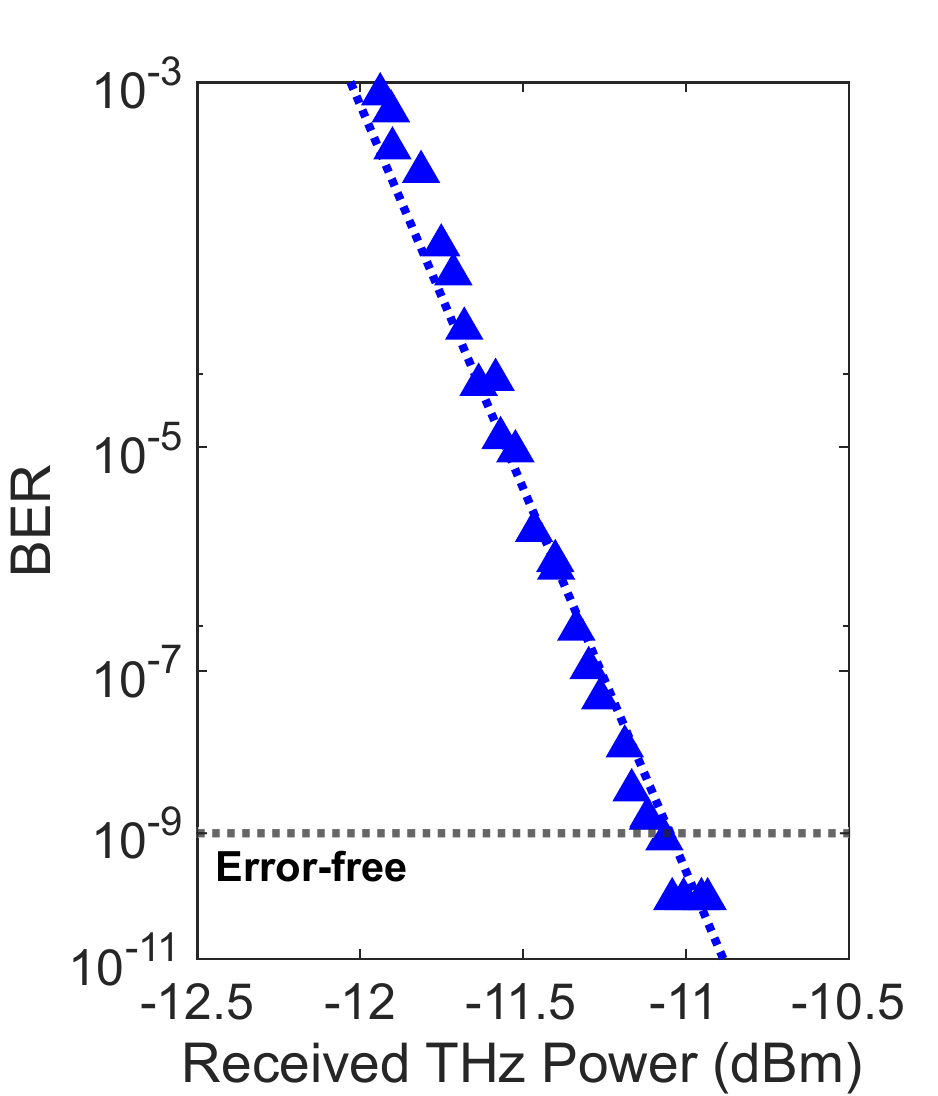}
  \caption{BER as a function of received THz power.}
  \label{fig:8b}
 \end{subfigure}
 \caption{BER.}
 \label{fig:8}
\end{figure}

Figure~\ref{fig:8}(a) shows the BER as a function of the UTC-PD input optical power, while Fig.~\ref{fig:8}(b) shows the BER as a function of the received THz power estimated from Fig.~\ref{fig:7}. Notably, error-free operation (BER~$<~1\times10^{-9}$) was successfully achieved. The UTC-PD input optical power threshold required for error-free operation was 14.92~dBm, and the corresponding received THz power threshold was $-11.05$~dBm. Since the UTC-PD output power is the limiting factor in the present setup, practical free-space deployment would require higher-gain antennas, THz amplifiers, and/or RF amplifiers to provide sufficient link margin.

\section{Discussion}\label{sec4}
\subsection{Free-space Link Scalability}

Although the transmission experiment in this study was conducted only in a back-to-back configuration, we also evaluated the feasibility of free-space THz wireless transmission using the Friis equation~\cite{Friis}. The received power $P_{\mathrm{Rx}}$ in dBm can be estimated as

\begin{equation}
\label{eq:friis-thz}
P_{\mathrm{Rx}}
= P_{\mathrm{Tx}}
+ G_{\mathrm{Tx}}
+ G_{\mathrm{Rx}}
+ 20\log_{10}\!\left(\frac{\lambda}{4\pi d}\right)
+ G_{\mathrm{THz}}
\end{equation}

where $P_{\mathrm{Tx}}$ and $P_{\mathrm{Rx}}$ are the transmit and receive powers, $G_{\mathrm{Tx}}$ and $G_{\mathrm{Rx}}$ are the gains of the transmit and receive antennas, $d$ is the transmission distance, $\lambda$ is the wavelength (1~mm at 300~GHz), and $G_{\mathrm{THz}}$ is the gain of a THz amplifier.

Using the maximum THz power generated in this work ($P_{\mathrm{Tx}}=-10.15~\mathrm{dBm}$) and assuming $G_{\mathrm{Tx}}=G_{\mathrm{Rx}}=50~\mathrm{dBi}$ and $G_{\mathrm{THz}}=15~\mathrm{dB}$, the received power exceeds the error-free threshold of $P_{\mathrm{Rx}}=-11.05~\mathrm{dBm}$ at transmission distances up to $d<49.64~\mathrm{m}$. Considering that THz amplifiers with gains exceeding these values have already been demonstrated~\cite{THz AMP 1,THz AMP 2}, these results suggest that the present approach could support error-free free-space THz transmission over several tens of meters when combined with high-gain antennas and THz amplification.

It should be noted, however, that the Friis equation assumes ideal free-space propagation. In practical scenarios, additional impairments such as multipath propagation, beam misalignment, absorption, and scattering may introduce substantial losses and degrade link performance. Moreover, near-field effects, corresponding to deviations from the ideal far-field conditions assumed by the Friis equation, cannot be neglected because realistic THz environments do not always satisfy far-field conditions~\cite{Near-field 1,Near-field 2}. These factors must therefore be carefully considered in future THz wireless systems based on photomixing and soliton microcomb sources.

\begin{table*}[htb]
  \centering
  \caption{Representative results of prior THz links using soliton-comb sources.}
  \label{tab:comparison_rep}
  \begin{tabular}{c c c c c c c}
    \hline
    Ref. & Carrier (GHz) & Modulation format & Data rate (Gbps) & Distance (m) & Receiver processing & BER \\
    \hline
    This work & 300 & OOK & 10 & Back-to-back & None & $<10^{-9}$ (no FEC) \\
    \cite{THz communication by soliton 1} & 560 & OOK & 2 & 0.6 & None & Q~=~3.40 (Near-FEC limit) \\
    \cite{THz communication by soliton 2} & 560 & OOK & 1 & 0.01 & None & Q~=~6.23 (Near error-free) \\
    \cite{THz communication by soliton 3} & 301 & 16QAM & 60 & 0.5 & Coherent DSP & EVM~=~11.9\% \\
    \cite{THz communication by soliton 4} & 300 & 64QAM & 0.0012 & 0.1 & Offline DSP + compensation & BER $<~4\times10^{-3}$ \\
    \cite{THz communication by soliton 5} & 302 & 32QAM & 250 & 55 & Offline DSP & BER $=2.66\times10^{-2}$ \\
    \cite{THz communication by soliton 6} & 436 & 16QAM & 100.78 & 10 & Offline DSP + FEC & Post-FEC error-free \\
    \hline
  \end{tabular}
\end{table*}

\subsection{Comparison with Previous Work and Future Outlook}
Table~\ref{tab:comparison_rep} summarizes representative THz transmission demonstrations based on soliton microcombs. Previous studies using IM-DD/OOK architectures achieved transmission at 2~Gbps over 0.6~m or 1~Gbps over 0.01~m, but did not achieve error-free performance, although near-error-free operation was reported~\cite{THz communication by soliton 1,THz communication by soliton 2}. These studies demonstrated the feasibility of simple THz links using soliton microcombs, but further improvements in transmission quality and link robustness were still necessary.

Several studies have also investigated coherent QAM transmission using soliton microcombs. Transmission at 60~Gbps using 16QAM modulation over 0.5~m was demonstrated, although BER was not reported~\cite{THz communication by soliton 3}. Another study enhanced UTC-PD output power by photomixing multiple comb lines and achieved low EVM in 64QAM transmission at a symbol rate of 200~kHz over 0.1~m~\cite{THz communication by soliton 4}. However, this approach required precise frequency-dispersion compensation and suffered from significant limitations in symbol rate. To the best of our knowledge, the most advanced demonstration to date achieved 32QAM transmission at 250~Gbps over 55~m, but relied on offline signal processing and did not achieve error-free operation~\cite{THz communication by soliton 5}. In addition, Ref.~\cite{THz communication by soliton 6} employed a dark-pulse Kerr comb to obtain higher comb-line power and demonstrated 16QAM transmission at 100.78~Gbps over 10~m, although offline processing and FEC were still required.

In contrast to many previous demonstrations based on coherent modulation formats and offline DSP, this work focuses on a simple IM-DD/OOK architecture and demonstrates FEC-free error-free transmission at 10~Gbps using microcomb-based photomixing. These results indicate the feasibility of robust low-BER THz transmission using soliton microcombs in low-complexity photonic-assisted wireless links.

Looking ahead, employing higher-power comb sources such as dark-pulse combs, together with high-gain antennas and THz amplifiers, is expected to further extend the transmission distance and capacity~\cite{Dark-pulse}. While the impact of phase noise is less critical in IM-DD OOK transmission than in coherent QAM systems, soliton microcombs remain attractive for future higher-order modulation formats owing to their mutual coherence and stability. In such phase-sensitive coherent transmission schemes, further advances in phase-noise stabilization and repetition-rate control are expected to improve transmission performance, particularly at low offset frequencies~\cite{Lower phase noise 1,Lower phase noise 2}. Together with progress in photonic integration and near-field-aware link design, these developments could support practical high-capacity THz photonic links for future fiber-wireless integrated systems.

\section{Conclusion}\label{sec5}
In this study, we generated a 300~GHz THz wave by photomixing two adjacent lines of a long-term-stabilized soliton comb and demonstrated 10~Gbps THz-band transmission in a simple IM-DD/OOK configuration. Although the transmission distance was limited to a short back-to-back waveguide link, the generated THz wave enabled error-free (BER~$<~1\times10^{-9}$) communication without FEC. These results demonstrate the feasibility of robust low-complexity THz photonic links based on soliton microcombs for short-range fiber-wireless integrated systems.

\section{Acknowledgments}
This work was supported by JST, CRONOS, Japan Grant Number JPMJCS24N7, and JSPS KAKENHI(JP24K17624). We thank Mr. H. Kumazaki for technical support. The first author gratefully acknowledges the financial support received from The Satomi Scholarship Foundation during the course of this research.

\end{document}